\documentstyle[multicol,psfig,prl,aps]{revtex}

\begin{document}

%%%\hfill Preprint BNL-NT-01/13\\

\title{
Transverse Momentum Spectra of $J/\psi$ and $\psi^{\prime}$ Mesons\\
from Quark Gluon Plasma Hadronization in Nuclear Collisions }

\vspace{0.3cm}

\author{\bf
K.A. Bugaev$^{a,b}$, M. Ga\'zdzicki$^{c}$ and M.I. Gorenstein$^{a,b}$ 
}

%\vspace*{0.4cm} 

\address{
\noindent
$^a$ Institut f\"ur Theoretische Physik, Universit\"at
Frankfurt, Germany\\
$^b$ Bogolyubov Institute for Theoretical Physics, Kiev, Ukraine
\\
$^c$ Institut f\"ur Kernphysik, Universit\"at Frankfurt, Frankfurt, Germany
}
\date{\today}

\maketitle

%\vspace{0.5cm}

\begin{abstract}
Recent results on transverse mass spectra of $J/\psi$ and $\psi^{\prime}$
mesons in central Pb+Pb collisions at 158 A$\cdot$GeV are considered. It is
shown that those results support a hypothesis of statistical production of
charmonia at hadronization and suggest the early thermal freeze--out of $%
J/\psi$ and $\psi^{\prime}$ mesons. Based on this approach the collective
transverse velocity of hadronizing quark gluon plasma is estimated to be $%
\langle v^H_T \rangle \approx 0.2$. Predictions for transverse mass spectra
of hidden and open charm mesons at SPS and RHIC are discussed.
\end{abstract}

\vspace*{0.5cm}

\noindent
\hspace*{2.cm}\begin{minipage}[t]{14.cm}
\end{minipage}

\vspace*{0.1cm}

%\pacs{ }

\vspace{0.5cm}

\begin{multicols}{2}

The idealized concepts of chemical (hadron multiplicities) and thermal
(hadron momentum distributions) freeze-outs were introduced to interpret
data on hadron production in relativistic nucleus--nucleus (A+A) collisions
\cite{Rev}. The first experimental results on yields and transverse mass ($%
m_T = \sqrt{p_T^2 + m^2}$) spectra suggested the following scenario: for the
most abundant hadron species ($\pi, N, K, \Lambda$) the chemical freeze-out,
which seems to coincide with the hadronization of the quark gluon plasma
(QGP), is followed by the thermal freeze--out occurring at a rather late
stage of the A+A reaction. In this letter we discuss whether the new data 
of NA50 \cite{Jpsi} on
transverse mass spectra of $J/\psi$ and $\psi^{\prime}$ mesons produced in
central Pb+Pb collisions at 158 A$\cdot$GeV are consistent with the above
picture. Our consideration is based on a recent hypothesis \cite{Ga:99} of
statistical production of charmonia at hadronization. We further suggest
that thermal freeze--out of charmonia coincides with the hadronization
transition. The consequences of this assumptions are in agreement with
existing data on $m_T$ spectra. They allow to extract the transverse flow
velocity of the hadronizing QGP and lead to new predictions which can be
tested by future measurements.

Rescattering among partons and, in the late stage of the reaction process,
hadrons created in relativistic A+A collisions should cause local
thermalization and the development of transverse collective flow of matter.
Thus the final transverse motion of hadrons can be considered as a
convolution of the transverse flow velocity of the freezing--out matter
element with the thermal motion of the hadrons in the rest frame of this
element. The resulting $m_T$ spectrum has approximately exponential shape:
\begin{equation}
\frac {1} {m_T} \cdot \frac {dn} {dm_T} \approx C \cdot e^{-m_T/T^*},
\end{equation}
where $C$ and $T^*$ are a normalization and an inverse slope parameters,
respectively. The $T^*$ parameter is related to the thermal freeze--out
temperature $T_f$ and the mean transverse flow velocity $\langle v_T^f
\rangle$. In the nonrelativistic approximation ($m_T<2m$) contributions from
thermal and collective particle motions can be separated and the inverse
slope parameter of the observed hadron spectrum can be expressed as \cite
{Si:99}:
\begin{equation}  \label{slope}
T^*~=~T_f~+~\frac{2}{\pi} \cdot m \cdot \langle v^f_T\rangle ^2~.
\end{equation}
Note that factor $2/\pi$ appears in Eq.(\ref{slope}) due to the cylindrical
geometry expected in high energy A+A collisions (see \cite{Si:99} for
details). In the case of the simultaneous thermal freeze--out of all hadrons
the inverse slope parameter $T^*$ should follow the linear dependence on $m$
given by Eq. (\ref{slope}). In Fig. 1 we present a compilation of inverse
slope parameters measured for various %light ($m < 2$ GeV/c$^2$)
hadron species in central Pb+Pb collisions at 158 A$\cdot$GeV. It is
observed that the simplistic hypothesis of common thermal freeze--out of all
hadrons is not supported by the data.

The freeze--out condition for pions can be determined independently of the
other hadron species from the results on two pion correlations. This
procedure leads to the following values of the pion thermal freeze--out
parameters \cite{Upline}:
\begin{equation}
T_{f}~=~120 \pm 12~{\rm {MeV}}~,~~~~\langle v_{T}^{f}\rangle ~=~0.5\pm 0.12~.
\label{T,v}
\end{equation}
In Fig. 1 the upper solid line indicates the dependence given by Eq.~(\ref
{slope}) for the pion freeze--out parameters. The values of $T^{\ast }$ for $%
K^{+}$ and $p$ are close to those given by this line (\ref{slope},\ref{T,v}%
). The results on $T^{\ast }$ for other hadrons are in general below the
'pion freeze-out line'. This may be\\ 

\vspace*{-1.5cm}

\begin{figure}
\hspace*{-1.4cm}\mbox{\psfig{figure=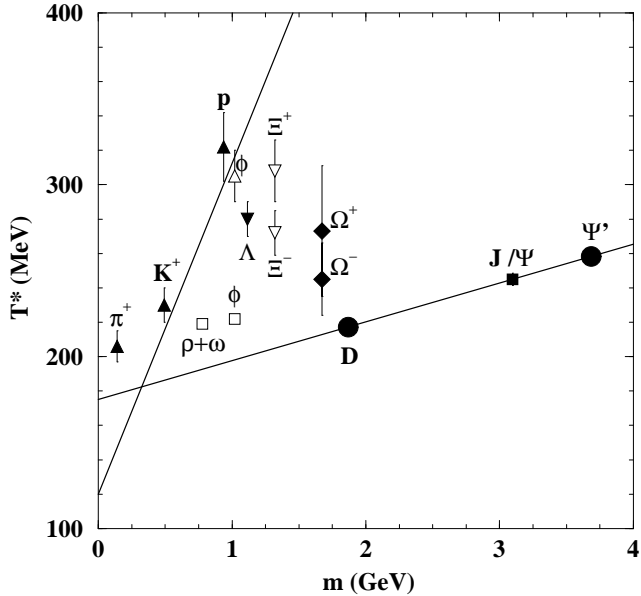,width=138mm}}

\vspace*{-0.5cm}

\noindent   
\caption{\label{fig:one}
The inverse slope parameter as a function of the particle mass for
central Pb+Pb collisions at 158 A$\cdot$GeV. The results for the following
hadrons are compiled from: $\pi$ mesons, $K$ mesons and protons 
[6] %\cite{trUF}
(closed triangles up), $\rho$ and $\omega$ mesons 
[7] %\cite{sqO} 
(open square), 
$\phi$ mesons [8] %\cite{trUO} 
(open triangles up) and 
[7] %\cite{sqO} 
(open square),
$\Lambda$ hyperons [9] % \cite{trDF} 
(closed triangle down), $\Xi$ hyperons
[10] %\cite{trDO} 
(open triangles down), $\Omega$ hyperons [11] %\cite{diam} 
(closed
diamonds), $J/\psi$ mesons [2] %\cite{Jpsi} 
(closed square).
The filled circles are predictions of the model for $\psi^\prime$ and D
mesons. The upper solid line corresponds to Eq.~(2) with the freeze--out
parameters (3) extracted from the two pion correlation data 
[5]. %\cite{Upline}.
The lower solid line is given by Eq.~(5) with parameters $T_H=175$~MeV and
$\langle v_T^H \rangle = 0.19$
which correspond to the QGP hadronization.
}
\end{figure}

%%%MYM

\noindent
interpreted in the following way. For
hadrons with ``low'' 
masses and large interaction cross sections the thermal
freeze--out takes place at the late stage of the expansion, i.e., at large $%
\langle v_{T}^{f}\rangle $ and small $T_{f}$ given by Eq.~(3). Heavy hadrons
(due to large mass and smaller cross section) are expected to decouple from
the system early thus leading to the smaller values of $T^{\ast }$ than
those expected from the 'pion freeze--out line' \footnote{%
For pions, in contrast to other hadrons, the relativistic effects neglected
in Eq. (\ref{slope}) are important. This is a main reason why the measured
pion inverse slope parameter is above the thermal pion freeze--out line in
Fig.~1.}. This is seen from the $m_{T}$-slopes of the hyperons and most
clearly can be illustrated by a small $T^{\ast }$ value of the $\Omega $
hyperon shown in Fig.~1. This explanation is in the line with the typical
hydrodynamical calculations (see, for instance, \cite{Dum:99}) -- the
inverse slope parameter $T^{\ast }$ of the hadron is smaller, if it
decouples early, i.e., at higher temperatures $T_{f}$, but smaller velocity $%
\langle v_{T}^{f}\rangle $. A more careful analysis requires the combination
of the hydrodynamic approach at the early stage of the expansion with
cascade model calculations at the latest stage \cite{BD,Sh}.

\vspace{0.3cm} The yields of produced light hadrons are surprisingly well
reproduced within the statistical approach to particle production. The
temperature parameter extracted from the fit to the multiplicity data is
found to be $T_H=175\pm$10~MeV \cite{Th:a,Th:b,Th:c}. This chemical
freeze--out temperature appears to be quite close to the expected
temperature of the transition between the hadron gas and the quark gluon
plasma. This fact suggests the possibility to ascribe the observed
statistical properties of hadron yield systematics at high energies to the
statistical nature of the hadronization process, i.e., the hadrons are
"born" at hadronization temperature $T_H$ into the state of chemical
equilibrium and their multiplicities are frozen--out afterwards.

Within this approach one can make a rough estimate of a lower limit for the
value of the measured inverse slope parameters:
\begin{equation}
T^*_{MIN} = T_H \approx 175~~ {\rm MeV}.
\end{equation}
This is done under two extreme assumptions: there is no collective
transverse flow of hadronizing matter and there is no rescattering between
produced hadrons. The measured values of $T^*$ parameter for all hadrons
satisfy the condition $T^* > T^*_{MIN}$ showing that the approach sketched
above leads to self-consistent results in the light hadron sector.

A very different approach is traditionally used in modeling the production
process of heavy hadrons like quarkonia. It is based on the assumption that
quarkonia are created significantly prior the hadronization of QGP. This in
general implies that the systematics of quarkonium production should be
different than the one established for light hadrons. However recently it
was found that multiplicity of the $J/\psi$ mesons in nuclear collisions at
high energies follows dependences well known from light hadron data \cite
{GaGo,Ga:98} and that it is consistent with the hypothesis of statistical
production of $J/\psi$ mesons at hadronization \cite{Ga:99}. In particular
the data on $J/\psi$ and $\psi^{\prime}$ yields in central Pb+Pb collisions
at 158 A$\cdot$GeV are consistent with the predictions of statistical models
\cite{Ga:99,Sh:97,ch1,ch2,ch3} for a typical values of $T_H \cong 175$~MeV
extracted from light hadron systematics.

The new hypothesis of statistical $J/\psi$ production at hadronization can
be further tested using data on $m_T$ spectra. As it follows from the above
discussion one expects for $J/\psi$ mesons an exponential shape of the $m_T$
distribution with the inverse slope parameter \mbox{$T^*(J/\psi) > T_H$.}
Recently
published experimental results of NA50 \cite{Jpsi} 
on transverse mass spectra of $J/\psi$ mesons
in central Pb+Pb collisions at 158 A$\cdot$GeV confirm this expectation: the
shape of the spectrum is approximately exponential with the fitted inverse
slope parameter $T^*(J/\psi) =245\pm 5$% is larger than $T_h$.
~MeV. The measured $T^*(J/\psi)$ value is significantly smaller than that
expected on the base of the pion freeze--out line ($T^* \cong$ 610 MeV).

The `low' value of $T^*$ for $J/\psi$ suggests its rather early thermal
freeze--out. One may argue that this is due to the large mass and low
interaction cross section of the $J/\psi$ meson. We postulate therefore that
the thermal freeze--out of $J/\psi$ coincides with the hadronization of QGP,
i.e., that the $J/\psi$ meson does not participate in the hadronic
rescattering after hadronization. It is, however, natural to expect that
there is a significant collective transverse flow of hadronizing QGP
developed at the early stage of partonic rescattering. Consequently the
inverse slope parameter of $J/\psi$ meson as well as all other hadrons for
which chemical and thermal freeze--outs coincide with hadronization can be
expressed as:
\begin{equation}  \label{slope1}
T^*_H~=~T_H~+~\frac{2}{\pi}~ \cdot m~ \cdot \langle v^H_T\rangle ^2~,
\end{equation}
where $\langle v^H_T\rangle$ is the mean transverse flow velocity of the QGP
at the hadronization. Assuming $T_H=175$~MeV and using measured value of $%
T^*(J/\psi) =245$ MeV we find from Eq.(\ref{slope1}): $\langle v^H_T\rangle
\cong 0.19$. As expected the obtained transverse flow velocity of QGP at
hadronization is significantly smaller than the transverse flow velocity of
pions ($\approx 0.5$). The linear $m$--dependence of $T^*_H$ (\ref{slope1})
is shown in Fig. 1 by the lower solid line. Within the approach discussed
here, Eq.~(\ref{slope1}) can be used to obtain a next estimate of the lower
limit of the measured inverse slope parameters for all hadrons. In fact the
values of the parameter $T^*$ for all light hadrons are higher than $T^*_H(m)
$. The recent results \cite{Jpsi} on 
the $m_T$ spectra of the $\psi^{\prime}$ meson
indicate that $T^*(\psi^{\prime}) \approx T^*_H(\psi^{\prime}) = 258 \pm 5$
MeV, which suggests that also the $\psi^{\prime}$ meson (like $J/\psi$) does
not participate in the hadronic rescattering.

One may expect that the thermal freeze--out may coincide with hadronization
also for the $D$ meson. Under this assumption we calculate the value of the
apparent temperature for the $D$ meson: $T^*(D) \cong T^*_H(D) \cong 217 $%
~MeV. Note that this result is significantly lower than the predictions of
the model of Ref. \cite{Lev:00a} although in \cite{Lev:00a} the same value
of the hadronization temperature was used.

Another important issue is the production of open and hidden charm particles
in A+A collisions at RHIC energies. One can expect stronger transverse
collective flow effects than at the SPS. This will lead to a linear mass
dependence (\ref{slope1}) of the apparent inverse slope with approximately
the same value of $T_H\cong 175$~MeV, but with a larger value of $\langle
v_{T}^{H}\rangle $. It is necessary to mention that a recent analysis \cite
{Redl:01} of the particle number ratios of the RHIC data leads to the above
value $T_H\cong 175$~MeV of the chemical freeze--out temperature. The
preliminary RHIC data of hadron $m_T$ spectra measured by STAR \cite{Star}
and PHENIX \cite{Phenix} collaborations support the fact of a larger value
of the transverse velocity. However, the measurements were made for $%
\pi^\pm, K^\pm, \bar{p},$ and $p$ only, and, therefore, this corresponds to
the late thermal freeze-out stage.

As long as the RHIC data on the inverse slope parameters for the open and
hidden charm mesons are absent it is interesting to compare the different
model predictions. Assuming $T_H\cong 175$~MeV we need the value of $\langle
v_T^H\rangle $ to estimate the inverse slope parameters for charmed hadrons
in our approach. The hydrodynamic calculations of Ref.~\cite{Dum:99} predict
the value of $\langle v_{T}^{H}\rangle \cong 0.30$ at the hadronization in
Au+Au collisions at RHIC. This leads to an increase of the inverse slopes of
charmed hadrons at RHIC in comparison to those values at SPS, e.g., $%
T^*(J/\psi) \cong T_H^*(J/\psi)\cong 350$~MeV and $T^*(D) \cong T^*_H(D)
\cong 280$~MeV.

Note that a significantly larger value of \mbox{$T^*(D) \cong 380$}~MeV in Au+Au
collisions at RHIC would be obtained, if the thermal freeze-out of D mesons
happens at temperature of $130$~MeV. On the other hand, the calculations
done with the PYTHIA generator for $p+p$ collisions show an even higher
value of the inverse slope for D mesons $T^*(D) \sim 440$ MeV \cite{Dum:99}.
In fact, the situation is even more uncertain as the dynamical transport
calculations of the HSD model (Hadron String Dynamics) made for RHIC energy
\cite{Cas:00} predict a rather small inverse slope of about 225~MeV for all
particle species. Therefore, the $T^*(m)$ systematics at RHIC for $m_T - m
\le 1$ GeV will provide an opportunity to find whether the statistical
hadronization works, whether a thermal freeze-out of charmed particles
happens simultaneously with hadronization and whether the transverse
collective flow at hadronization of the QGP is stronger at RHIC than at SPS.

In summary, recent results on transverse mass spectra of $J/\psi$ and $%
\psi^{\prime}$ mesons in central Pb+Pb collisions at 158 A$\cdot$GeV are
considered. It is shown that these data support the hypothesis of the
statistical production of charmonia at hadronization and suggest a
simultaneous  hadronization and the thermal freeze--out for $J/\psi$ and $%
\psi^{\prime}$ mesons. Based on this approach the collective transverse
velocity of hadronizing quark gluon plasma is estimated to be $\langle v^h_T
\rangle \approx 0.2 $. Prediction for transverse mass spectra of hidden and
open charm mesons at SPS and RHIC are discussed.

\vspace*{0.4cm}

\begin{center}
{\bf Acknowledgments}
\end{center}

\vspace*{0.2cm}

The authors are thankful to S. V. Akkelin,
A. Dumitru, R. Pisarski, J. Schaffner-Bielich
and Yu. Sinyukov for stimulating discussions. 
The very fruitful and stimulating discussions
with L.D. McLerran are  appreciated.
K.A.B. gratefully acknowledges the warm hospitality of the 
BNL
Nuclear Theory Group,
where parts of this work were done.
The financial support of 
DAAD, Germany, is acknowledged. The research described in this
publication was made possible in part by Award No. UP1-2119 of the U.S.
Civilian Research \& Development Foundation for the Independent States of
the Former Soviet Union (CRDF).
This manuscript has been authorized under 
Contract No. DE-AC02-98CH10886 with the U.S. Department of Energy.

%\begin{references}

\end{multicols}


\begin{thebibliography}{99}

\vspace{-0.9cm}

\bibitem{Rev}  U. A.~Wiedemann and U. Heinz, Phys. Rep. {\bf 319} 145 (1999)
and references therein.

\bibitem{Jpsi}  M.C. Abreu {\it et al.} (NA50 Collaboration), Phys. Lett.
{\bf B499} 85 (2001).

\bibitem{Ga:99}  M. Ga\'{z}dzicki and M.I. Gorenstein, Phys. Rev. Lett. {\bf %
83} 4009 (1999).

\bibitem{Si:99}  Yu. M. Sinyukov, S.V. Akkelin and N.Xu, Phys. Rev. {\bf C59}
3437 (1999).

\bibitem{Upline}  H. Appelshauser {\it et. al.} (NA49 Collab.), Eur. Phys.
J. {\bf C2} 661 (1998).

\bibitem{trUF}  S. V. Afanasjev {\it et al.} (NA49 Collab.), Phys. Lett.{\bf %
B486} 22 (2000).

\bibitem{sqO}  M. C. Abreu {\it et al.} (NA50 Collab.), Nucl. Phys. {\bf A661%
} 534c (1999).

\bibitem{trUO}  S. V. Afanasjev {\it et al.} (NA49 Collab.), Phys. Lett.
{\bf B491} 59 (2000).

\bibitem{trDF}  S. Margetis {\it et al.} (NA49 Collab.), J. Phys. {\bf G25}
189 (1999).

\bibitem{trDO}  R. A. Barton {\it et al.} (NA49 Collab.), J. Phys. {\bf G27}
367 (2001).

\bibitem{diam}  W. Beusch {\it et al.} (WA97 Collab.), J. Phys. {\bf G27}
375 (2001).

\bibitem{Dum:99}  A. Dumitru and C. Spieles, Phys. Lett. {\bf B446} 326
(1999).

\bibitem{BD}  S. A. Bass and A. Dumitru, Phys. Rev. {\bf C61} 064909 (2000).

\bibitem{Sh}  D. Teaney, J. Lauret and E.V. Shuryak, {\bf nucl-th/0104041}
(2001).

\bibitem{Th:a}  G. D.~Yen and M.I.~Gorenstein, Phys. Rev. {\bf C59} 2788
(1999).

\bibitem{Th:b}  P. Braun-Munzinger, I. Heppe and J. Stachel, Phys. Lett.
{\bf B465} 15 (1999).

\bibitem{Th:c}  J.~Cleymans and K. Redlich, Phys. Rev. {\bf C60} 054908
(1999); F. Becattini {\it et al.}, {\bf hep-ph/0002267} (2000).

\bibitem{GaGo}  M. Ga\'zdzicki and M. I. Gorenstein, Acta Phys. Polon. {\bf %
B30} 2705 (1999).

\bibitem{Ga:98}  M. Ga\'zdzicki, Phys. Rev. {\bf C60} 054903 (1999).

\bibitem{Sh:97}  H. Sorge, E. Shuryak and I. Zahed Phys. Rev. Lett. {\bf 79}
2775 (1997).

\bibitem{ch1}  P.~Braun-Munzinger and J.~Stachel, Phys. Lett. {\bf B490},
196 (2000); Nucl.Phys. {\bf A690} 119 (2001).

\bibitem{ch2}  M.~I.~Gorenstein, A.~P.~Kostyuk, H.~St\"{o}cker and
W.~Greiner, {\bf hep-ph/0010148} (Phys. Lett. B, in print); {\bf %
hep-ph/0012015} (J. Phys. G, in print); {\bf hep-ph/0104071}.

\bibitem{ch3}  M.~I.~Gorenstein, A.~P.~Kostyuk, L.~McLerran, H.~St\"{o}cker
and W.~Greiner, Preprint {\bf hep-ph/0012292} (2000).

\bibitem{Lev:00a}  P. Csizmadia and P. Levai, Preprint {\bf hep-ph/0008195}
(2000).

\bibitem{Redl:01}  P. Braun-Munzinger {\it et al.}, Preprint {\bf %
hep-ph/0105229} (2001); \hfill\\ 
W. Florkowski, W. Broniowski and M. Michalec,
Preprint {\bf nucl-th/0106009} (2001).


\bibitem{Star}  N.~Xu and M.~Kaneta, Preprint {\bf nucl-ex/0104021} (2001).

\bibitem{Phenix}  J.~Velkovska, for the PHENIX Collaboration, Preprint {\bf %
nucl-ex/0105012} (2001).

\bibitem{Cas:00}  W. Cassing, E. L. Bratkovskaya and A. Sibirtsev, {\bf %
nucl-th/0010071} (2000).

\end{thebibliography}
\end{document}